\documentclass[11pt, a4paper]{article}
\usepackage[margin=1in]{geometry} 
\usepackage{amsmath,amsfonts}
\usepackage{algorithmic}
\usepackage{algorithm}
\usepackage{array}
\usepackage[caption=false,font=normalsize,labelfont=sf,textfont=sf]{subfig}
\usepackage[colorlinks=true,linkcolor=blue,citecolor=blue]{hyperref}
\usepackage{orcidlink}
\usepackage{textcomp}
\usepackage{url}
\usepackage{verbatim}
\usepackage{graphicx}
\usepackage{cite}
\usepackage{booktabs}
\usepackage{multirow}
\usepackage{bm}
\usepackage{float}
\usepackage[utf8]{inputenc}
\usepackage{authblk} 

\newcommand{\figsref}[2]{Fig.~\hyperref[#1]{\ref*{#1}#2}}
\newcommand{\figref}[2]{Fig.~\hyperref[#1]{\ref*{#1}(#2)}}
\newcommand{\justsub}[2]{\hyperref[#1]{(#2)}}
\hypersetup{
    colorlinks=true,   
    urlcolor=blue,     
    linkcolor=blue,   
    citecolor=blue     
}

\begin{document}

\title{A Conditional Denoising Diffusion Probabilistic Model for RFI Mitigation in Synthetic Aperture Interferometric Radiometer}

\author[1]{Yuankai Luo\orcidlink{0009-0008-5890-0286}}
\author[1]{Han Zhou\orcidlink{0009-0003-4631-7150}}
\author[1]{Jinlong Hao\orcidlink{0009-0004-5634-490X}}
\author[2]{Dong Zhu\orcidlink{0000-0001-6423-5050}}
\author[2]{Fei Hu\orcidlink{0000-0003-3625-286X}\thanks{Corresponding author: Fei Hu (e-mail: hufei@hust.edu.cn). This work was supported in part by the National Natural Science Foundation of China under Grant 62271219; and in part by the Fundamental Research Founds for the Central Universities (Huazhong University of Science and Technology), China.}}

\affil[1]{\small School of Electronic Information and Communications, Huazhong University of Science and Technology, Wuhan 430074, China}
\affil[2]{\small School of Electronic Information and Communications and the National Key Laboratory of Science and Technology on Multi-Spectral Information Processing, Huazhong University of Science and Technology, Wuhan 430074, China}

\date{} 

\maketitle

\footnotetext{\copyright~2024 IEEE. Personal use of this material is permitted. Permission from IEEE must be obtained for all other uses, in any current or future media, including reprinting/republishing this material for advertising or promotional purposes, creating new collective works, for resale or redistribution to servers or lists, or reuse of any copyrighted component of this work in other works.}

\begin{abstract}
In Earth remote sensing, spatial-frequency domain visibility samples are inversely transformed into spatial-domain brightness temperature (BT) images through the signal processing pipeline of synthetic aperture interferometric radiometers (SAIR). However, L-band radio-frequency interference (RFI) contaminates the measured visibilities and severely degrades BT image quality, thereby impairing geophysical parameter retrieval. To address this issue, we propose VFDM, a Visibility-Function Diffusion Model based on Denoising Diffusion Probabilistic Models (DDPM), to mitigate RFI in the spatial-frequency domain while preserving fine-scale structures consistent with natural scene statistics. Furthermore, we construct a comprehensive dataset comprising more than ten thousand pairs of RFI-free natural scene visibility sample sets and their corresponding simulated contaminated counterparts, categorized by varying RFI intensities, numbers, and distributions. Finally, comprehensive experiments on both simulated and real-world data demonstrate the effectiveness and robustness of the proposed VFDM-based approach.
\end{abstract}

\vspace{1em}

\section{Introduction}
Synthetic aperture interferometric radiometers (SAIRs) enable high-resolution passive microwave imaging by synthetically extending the effective aperture through interferometric measurements. Originally developed in radio astronomy, this principle has been successfully adopted in Earth observation missions. A representative example is the MIRAS instrument onboard ESA’s Soil Moisture and Ocean Salinity (SMOS) satellite, launched in 2009, which retrieves global soil moisture and ocean salinity from brightness temperature (BT) observations \cite{BTobservation2}.

Owing to its high sensitivity, MIRAS is vulnerable to radio-frequency interference (RFI) from radars and communication systems \cite{satellites1}, even though it operates in the protected L-band (1400--1427~MHz). Moreover, the band-limited sampling inherent to SAIRs exacerbates RFI contamination, resulting in severe artifacts in reconstructed BT images and degraded geophysical retrieval accuracy.

Existing RFI mitigation methods for SMOS can be broadly categorized into three classes.
Array factor–based approaches suppress RFI by adaptively reweighting the $u$–$v$ baselines \cite{baseline2}, but their effectiveness relies on accurate prior knowledge of interference locations and power levels, which is often unavailable.
Filtering-based methods exploit the spectral or temporal characteristics of RFI using designed filters \cite{filter2}; however, they may suppress natural signals and typically assume simplified interference models.
Point-source cancellation techniques explicitly detect and subtract RFI components \cite{component1,component2}, yet their performance critically depends on precise estimation of the number, locations, and strengths of RFI sources, which is challenging in low signal-to-interference or multi-emitter scenarios.

Overall, most existing methods strongly depend on prior information and struggle to mitigate RFI while preserving fine-scale structures without introducing artifacts. To overcome these limitations, we propose VFDM, a Visibility-Function Diffusion Model for RFI mitigation. By leveraging the generative capability of diffusion models, VFDM recovers fine-grained details consistent with natural scene statistics, conditioning directly on contaminated visibility measurements in the spatial-frequency domain. Extensive experiments under diverse RFI conditions, including hybrid scenarios with multiple sources of varying intensities, demonstrate that VFDM effectively suppresses RFI-contaminated visibility components while faithfully reconstructing the underlying natural scene. To the best of our knowledge, this is the first work to apply diffusion models to RFI mitigation in synthetic aperture interferometric radiometry.

\section{Preliminaries} 
\label{section:Preliminaries}
\subsection{Principle of SAIR}

SAIR measures cross-correlations between signals received at different baselines using a complex correlator, yielding spatial-frequency domain measurements known as visibility function samples \cite{correlator}. Assuming negligible spatial decorrelation, the relationship between the visibility function and the brightness temperature (BT) image can be expressed as a spatial integral:
\begin{equation}
\begin{split}
V(u, v) &= \iint_{\xi^2 + \eta^2 \le 1}
\frac{k Z}{\lambda_c^2}
\frac{T_B(\xi, \eta)\, \lvert f(\xi, \eta)\rvert^2}
{\sqrt{1 - \xi^2 - \eta^2}} \\
&\quad \times
e^{-j 2\pi (u \xi + v \eta)}
\, \mathrm{d}\xi \, \mathrm{d}\eta
\end{split}
\label{1}
\end{equation}
where $\xi$ and $\eta$ are direction cosine coordinates; $k$ is the Boltzmann constant; $Z$ denotes the air impedance; $f(\xi,\eta)$ is the normalized antenna field pattern; and $u$ and $v$ are baselines normalized by the center wavelength $\lambda_c$. Following \cite{modified}, the modified brightness temperature is defined as
\begin{equation}
T_M(\xi,\eta)=
\frac{k Z}{\lambda_c^{2}}
\,
\frac{T_B(\xi,\eta)\,\lvert f(\xi,\eta)\rvert^{2}}
{\sqrt{1-\xi^{2}-\eta^{2}}}
\label{2}
\end{equation}
Substituting \eqref{2} into \eqref{1}, the visibility function reduces to a 2D Fourier transform:
\begin{equation}
V(u,v)=
\iint_{\xi^{2}+\eta^{2}\le 1}
T_M(\xi,\eta)\,
e^{-j2\pi\,(u\xi+v\eta)}
 \mathrm{d}\xi\, \mathrm{d}\eta.
\label{3}
\end{equation}
Accordingly, the modified BT image can be reconstructed via the inverse Fourier transform:
\begin{equation}
T_M(\xi,\eta)=
\iint
V(u,v)\,
e^{j2\pi\,(u\xi+v\eta)}
\, \mathrm{d}u\, \mathrm{d}v.
\label{4}
\end{equation}
With discrete and band-limited sampling in the $u$--$v$ plane, this inversion is implemented using the Inverse Discrete Fourier Transform (IDFT):
\begin{equation}
\hat{T}_M(\xi,\eta)=
\Delta s
\sum_{i=1}^{N_v}
V(u_i,v_i)\,
e^{j2\pi\,(u_i\xi+v_i\eta)}.
\label{5}
\end{equation}

In particular, the signal model of SAIR can be interpreted from a general array signal processing perspective. Consider an SAIR array with $m$ antennas observing $l$ RFI sources ($l<m$). Let $\mathbf{a}_i$ and $s_i$ denote the steering vector and spatial signal of the $i$th source, respectively. Including measurement noise $\mathbf{n}$, the RFI-plus-noise received vector is modeled as
\begin{equation}
\mathbf{y} = \sum_{i=1}^{l} \mathbf{a}_i s_i + \mathbf{n}.
\label{6}
\end{equation}
In interferometric radiometry, the second-order statistics of the received signal vector $\mathbf{y}$ are characterized by the covariance matrix obtained via correlation and time averaging. Specifically, the RFI covariance matrix $\mathbf{R}_I$ is modeled as:
\begin{equation}
\mathbf{R}_I = \langle \mathbf{y}\mathbf{y}^H \rangle
= \sum_{i=1}^{l} \alpha_i^2 \mathbf{a}_i \mathbf{a}_i^H + \mathbf{N},
\label{7}
\end{equation}
where $\langle\cdot\rangle$ denotes the statistical expectation.

From a subspace perspective \cite{perspective}, the total measured covariance matrix $\mathbf{R}$ implies a superposition of signals. It decomposes into the RFI component $\mathbf{R}_I$ and the natural scene component $\mathbf{R}_S$, as defined by:
\begin{equation}
\mathbf{R} = \mathbf{R}_I + \mathbf{R}_S.
\label{8}
\end{equation}
Moreover, this statistical representation is physically linked to the imaging domain. Each element of $\mathbf{R}$ corresponds to a sampled visibility at the associated baseline $(u_{ij}, v_{ij})$ \cite{in}, as expressed in:
\begin{equation}
[\mathbf{R}]_{ij} = V(u_{ij}, v_{ij}).
\label{9}
\end{equation}

\subsection{DDPM}
\label{section:DDPM}
DDPMs \cite{DDPM} established diffusion-based frameworks as powerful competitors to GANs for high-quality image generation. These latent variable models learn to map a data distribution $p_\theta(\mathbf{x}_0)$ by reversing a gradual Gaussian noise process. The standard implementation utilizes a U-Net as the underlying neural backbone to predict the added noise at each timestep.

Formally, as defined in \cite{to}, we consider the latent variables $\mathbf{x}_1, \dots, \mathbf{x}_T$ having the same dimensionality as the data $\mathbf{x}_0 \sim q(\mathbf{x}_0)$. The model is defined by a reverse process $p_\theta(\mathbf{x}_{0:T})$, which is a Markov chain with learned Gaussian transitions starting from 
a standard Gaussian prior 
$p(\mathbf{x}_T) = \mathcal{N}(\mathbf{x}_T; \mathbf{0}, \mathbf{I})$:
\begin{equation}
p_\theta(\mathbf{x}_{0:T})
=
p(\mathbf{x}_T)
\prod_{t=1}^{T}
p_\theta(\mathbf{x}_{t-1} \mid \mathbf{x}_t).
\label{eq:reverse_process}
\end{equation}

\begin{equation}
p_\theta(\mathbf{x}_{t-1} \mid \mathbf{x}_t)
=
\mathcal{N}
\!\left(
\mathbf{x}_{t-1};
\boldsymbol{\mu}_\theta(\mathbf{x}_t, t),
\mathbf{\Sigma}_\theta(\mathbf{x}_t, t)
\right).
\label{eq:reverse_transition}
\end{equation}

The fixed approximate posterior $q(\mathbf{x}_{1:T}|\mathbf{x}_0)$ is known as the forward process or the diffusion process. This process is a fixed Markov chain that gradually injects Gaussian noise into the data according to a pre-defined variance schedule 
\begin{equation}
q(\mathbf{x}_{1:T} \mid \mathbf{x}_0)
=
\prod_{t=1}^{T}
q(\mathbf{x}_t \mid \mathbf{x}_{t-1}),
\label{eq:forward_process}
\end{equation}

\begin{equation}
q(\mathbf{x}_t \mid \mathbf{x}_{t-1})
=
\mathcal{N}
\!\left(
\mathbf{x}_t;
\sqrt{1-\beta_t}\,\mathbf{x}_{t-1},
\beta_t \mathbf{I}
\right).
\label{eq:forward_transition}
\end{equation}

\begin{figure*}[t]
    \centering
    \includegraphics[width=1 \textwidth]{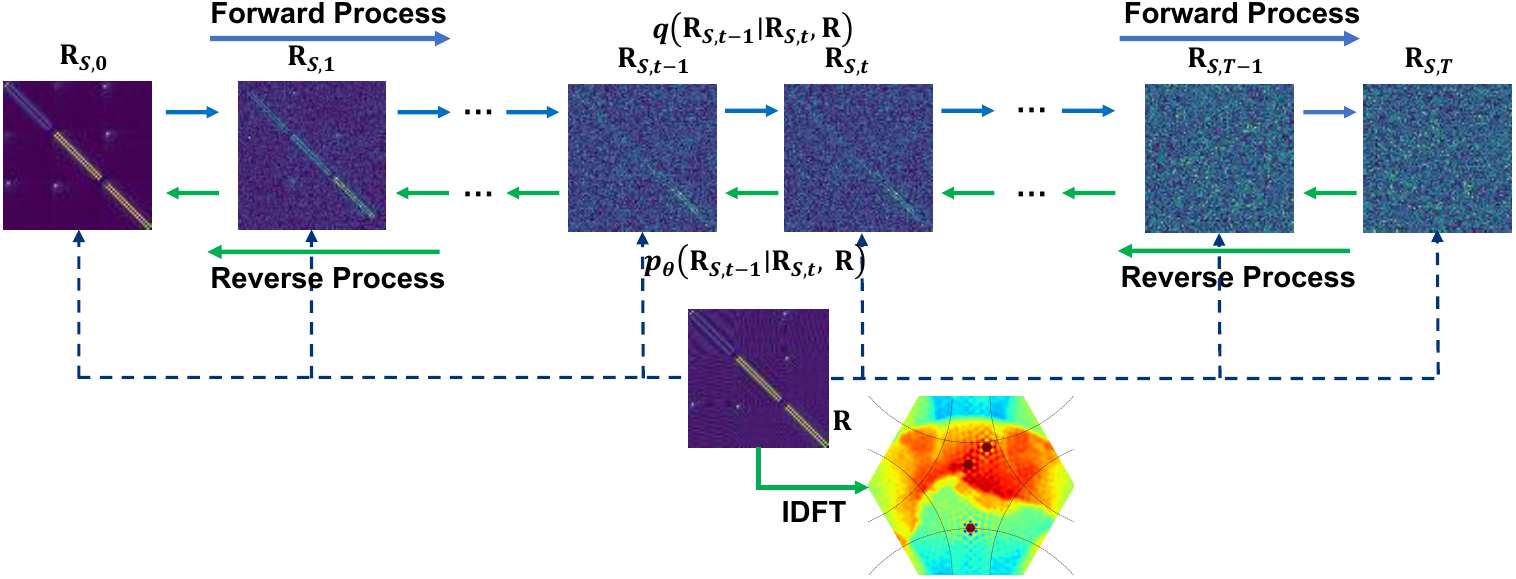}
    \caption{Overview of the Proposed VFDM. The diffusion process starts from $\mathbf{R}_{S,0}$ and Gaussian noise is gradually added to $\mathbf{R}_{S,t-1}$ to obtain $\mathbf{R}_{S,t}$ until $t = T$. The reverse process starts from random noise and generates $\mathbf{R}_{S,t-1}$ from $\mathbf{R}_{S,t}$. The contaminated covariance matrix is incorporated into the reverse process as a condition to guide the reconstruction of the natural covariance matrix.}
    \label{fig:vfdm}
\end{figure*}

\section{RFI Mitigation Based on Denoising Diffusion Probabilistic Model}
Suppose $\mathbf{R} \in \mathbb{C}^{m \times m}$ represents an acquired dirty visibility sample, which is decomposed into two components: the natural scene component $\mathbf{R}_S$ and the RFI component $\mathbf{R}_I$. The objective of RFI mitigation is to estimate the posterior distribution $q(\mathbf{R}_S \mid \mathbf{R})$. To effectively model this distribution and recover the underlying natural visibility information, we formulate the diffusion process as follows:
\begin{equation}
p_\theta(\mathbf{R}_S \mid \mathbf{R})
= \int p_\theta(\mathbf{R}_{S,0:T} \mid \mathbf{R}) \,\mathrm{d}\mathbf{R}_{S,1:T}.
\label{10}
\end{equation}
The reverse process $p_\theta(\mathbf{R}_{S,0:T} \mid \mathbf{R})$ is defined as:
\begin{equation}
p_\theta(\mathbf{R}_{S,0:T} \mid \mathbf{R})
= p(\mathbf{R}_{S,T})
\prod_{t=1}^{T}
p_\theta\bigl(\mathbf{R}_{S,t-1} \mid \mathbf{R}_{S,t}, \mathbf{R}\bigr),
\label{11}
\end{equation}
where each reverse transition follows a Gaussian distribution:
\begin{equation}
p_\theta\bigl(\mathbf{R}_{S,t-1} \mid \mathbf{R}_{S,t}, \mathbf{R}\bigr)
=
\mathcal{N}\!\left(
\mathbf{R}_{S,t-1};
\mu_\theta(\mathbf{R}_{S,t}, t, \mathbf{R}),
\sigma_t^2 \mathbf{I}
\right).
\label{12}
\end{equation}
The forward diffusion process is defined as a Markov chain similar to that in the original DDPM \cite{DDPM}:
\begin{equation}
q\!\left(\mathbf{R}_{S,1:T}\mid \mathbf{R}_{S,0}, \mathbf{R}\right)
:=\prod_{t=1}^{T} q\!\left(\mathbf{R}_{S,t}\mid \mathbf{R}_{S,t-1}, \mathbf{R}\right),
\label{13}
\end{equation}
\begin{equation}
q\!\left(\mathbf{R}_{S,t}\mid \mathbf{R}_{S,t-1}, \mathbf{R}\right)
:=\mathcal{N}\!\left(\mathbf{R}_{S,t};\, \alpha_t \mathbf{R}_{S,t-1},\, \beta_t^{2}\mathbf{I}\right).
\label{14}
\end{equation}
This process progressively injects Gaussian noise according to the schedules $\{\alpha_t\}$ and $\{\beta_t\}$.

Following recent work \cite{work}, we adopt a modified noise schedule. Define $\bar{\alpha}_t=\prod_{i=1}^{t}\alpha_i$, $\bar{\beta}_t^{2}=\sum_{i=1}^{t}\frac{\bar{\alpha}_t^{2}}{\bar{\alpha}_i^{2}}\beta_i^{2}$, with $\bar{\alpha}_0=1$ and $\bar{\beta}_0=0$. Then,
\begin{equation}
q\!\left(\mathbf{R}_{S,t}\mid \mathbf{R}_{S,0}, \mathbf{R}\right)
=\mathcal{N}\!\left(\mathbf{R}_{S,t};\, \bar{\alpha}_t \mathbf{R}_{S,0},\, \bar{\beta}_t^{2}\mathbf{I}\right),
\label{15}
\end{equation}
\begin{equation}
\begin{aligned}
& q\!\left(\mathbf{R}_{S,t-1}\mid \mathbf{R}_{S,t}, \mathbf{R}_{S,0}, \mathbf{R}\right) \\
&= \mathcal{N}\!\left(
\mathbf{R}_{S,t-1};\,
\tilde{\mu}_t(\mathbf{R}_{S,t},\mathbf{R}_{S,0}),\,
\tilde{{\beta}_t^{2}}\mathbf{I}
\right).
\end{aligned}
\label{16}
\end{equation}
where
$\tilde{\mu}_t(\mathbf{R}_{S,t},\mathbf{R}_{S,0})
=\frac{\alpha_t \bar{\beta}_{t-1}^{2}}{\bar{\beta}_t^{2}}\mathbf{R}_{S,t}
+\frac{\bar{\alpha}_{t-1}\beta_t^{2}}{\bar{\beta}_t^{2}}\mathbf{R}_{S,0}$,
$\tilde{\beta}_t^{2}=\frac{\beta_t^{2}\bar{\beta}_{t-1}^{2}}{\bar{\beta}_t^{2}}$.
With this schedule, $\bar{\alpha}_T \approx 0$. The model is trained by minimizing the variational bound on the negative log-likelihood, following the DDPM framework \cite{DDPM}:

\begin{equation}
\begin{aligned}
& \mathbb{E}\big[ - \log p_\theta ( \mathbf{R}_{S,0} | \mathbf{R} ) \big] \\
& \le \mathbb{E}_q \bigg[
    - \log p ( \mathbf{R}_{S,T} | \mathbf{R} ) \\
& \qquad 
    - \sum_{t \ge 1} \log \frac{p_\theta ( \mathbf{R}_{S,t-1} | \mathbf{R}_{S,t}, \mathbf{R} )}{q ( \mathbf{R}_{S,t} | \mathbf{R}_{S,t-1}, \mathbf{R} )}
\bigg] =: \mathcal{L}.
\end{aligned}
\label{17}
\end{equation}
We adopt the parameterization:
\begin{equation}
\mu_\theta(\mathbf{R}_{S,t},t,\mathbf{R})
=\frac{1}{\alpha_t}\left(
\mathbf{R}_{S,t}
-\frac{\beta_t^{2}}{\bar{\beta}_t}\,\bm{\epsilon}_\theta(\mathbf{R}_{S,t},t,\mathbf{R})
\right).
\label{18}
\end{equation}
Here, $\bm{\epsilon}_\theta$ predicts the noise component from $\mathbf{R}_{S,t}$. For $t=0$, $\mathbf{R}_{S,0}$ is directly output; otherwise,
\begin{equation}
\mathbf{R}_{S,t-1}
=
\frac{1}{\alpha_t}\left(
\mathbf{R}_{S,t}
-\frac{\beta_t^{2}}{\bar{\beta}_t}\,\bm{\epsilon}_\theta(\mathbf{R}_{S,t},t,\mathbf{R})
\right)
+\sigma_t \mathbf{z}_t,
\label{19}
\end{equation}
where $\mathbf{z}_t\sim\mathcal{N}(0,\mathbf{I})$.
Based on Eq.~\eqref{16} and the parameterization in Eq.~\eqref{18}, the loss can be simplified as
\begin{equation}
L_{\text{simple}}(\theta)
=\mathbb{E}_{\mathbf{R}_{S,0},t,\bm{\epsilon}}\left\|
\bm{\epsilon}-\bm{\epsilon}_\theta(\bar{\alpha}_t\mathbf{R}_{S,0}+\bar{\beta}_t\bm{\epsilon},\,t,\,\mathbf{R})
\right\|_2^{2}
\label{20}
\end{equation}
where $\bm{\epsilon}\sim\mathcal{N}(0,\mathbf{I})$ and $t\sim \text{Uniform}(0,\ldots,T)$.

\textbf{Training:}
At each iteration, paired samples $\{\mathbf{R}_S,\mathbf{R}\}\sim q(\mathbf{R}_S,\mathbf{R})$ are drawn from the training set. A noise vector $\bm{\epsilon}\sim\mathcal{N}(\mathbf{0},\mathbf{I})$ and a diffusion step $t\sim\mathrm{Uniform}(\{1,\ldots,T\})$ are randomly sampled. Consistent with Eq.~\eqref{15}, the noisy input is constructed as $\mathbf{R}_{S,t}=\bar{\alpha}_t\mathbf{R}_{S,0}+\bar{\beta}_t\bm{\epsilon}$. The network is trained by minimizing the loss $L(\theta)=\|\bm{\epsilon}-\bm{\epsilon}_\theta(\mathbf{R}_{S,t},t,\mathbf{R})\|_2^2$ using gradient descent until convergence. The network is implemented using a U-Net architecture. Model parameters are optimized using the Adam optimizer with an initial learning rate of $6 \times 10^{-4}$, which is decayed by a factor of 0.5 following a cosine annealing schedule. The model is trained for a total of 80k steps.

\textbf{Sampling:}
Given an observation $\mathbf{R}$, the sampling process starts from Gaussian noise $\mathbf{R}_{S,T}\sim\mathcal{N}(\mathbf{0},\mathbf{I})$. For $t=T,\ldots,1$, an estimate of the natural covariance is first computed by inverting the forward process: $\hat{\mathbf{R}}_{S,0} =\frac{1}{\bar{\alpha}_t}\!\left( \mathbf{R}_{S,t}-\bar{\beta}_t\bm{\epsilon}_\theta(\mathbf{R}_{S,t},t,\mathbf{R}) \right)$. Then, the generalized update step reconstructs the previous state: $\mathbf{R}_{S,t-1} =\bar{\alpha}_{t-1}\hat{\mathbf{R}}_{S,0} +\sqrt{\bar{\beta}_{t-1}^2-\sigma_t^2}\,\bm{\epsilon}_\theta(\mathbf{R}_{S,t},t,\mathbf{R}) + \sigma_t \mathbf{z}_t$, where $\mathbf{z}_t \sim \mathcal{N}(\mathbf{0}, \mathbf{I})$ is standard Gaussian noise. The variance term $\sigma_t$ depends on the hyperparameter $\eta$: $\sigma_t = \eta \tilde{\beta}_t = \eta \sqrt{ \beta_t^2 \bar{\beta}_{t-1}^2 / \bar{\beta}_t^2 }$. After completing all steps, $\mathbf{R}_{S,0}$ is returned as the reconstructed natural covariance through VFDM.

In summary, the diffusion process and the reverse process in VFDM are illustrated in Fig.~\ref{fig:vfdm}.

\section{Dataset and Results}
\label{section:result}
We constructed a specialized dataset comprising 13,007 sample pairs to facilitate model training and evaluation. Specifically, a randomized RFI simulator \cite{following} was applied to SMOS L1A observations to generate entries pairing simulated contaminated visibility samples with their ground-truth natural counterparts. These test scenarios encompass diverse conditions, featuring 1 to 8 RFI sources with intensities spanning weak, medium, strong, very strong, and hybrid regimes.

Utilizing this dataset, we compare VFDM against three representative methods—CLEAN \cite{CLEAN}, RPCA \cite{RPCA}, and RNN-DFT \cite{perspective}—within the alias-free field of view (AF-FOV). Performance is evaluated using Root Mean Square Error (RMSE) and Structural Similarity Index Measure (SSIM) for simulated data, while the proposed Total Reconstruction Error (TRE) is used to validate real SMOS observations in the absence of reference data.

\begin{figure*}[t] 
    \centering
    \includegraphics[width=1 \textwidth]{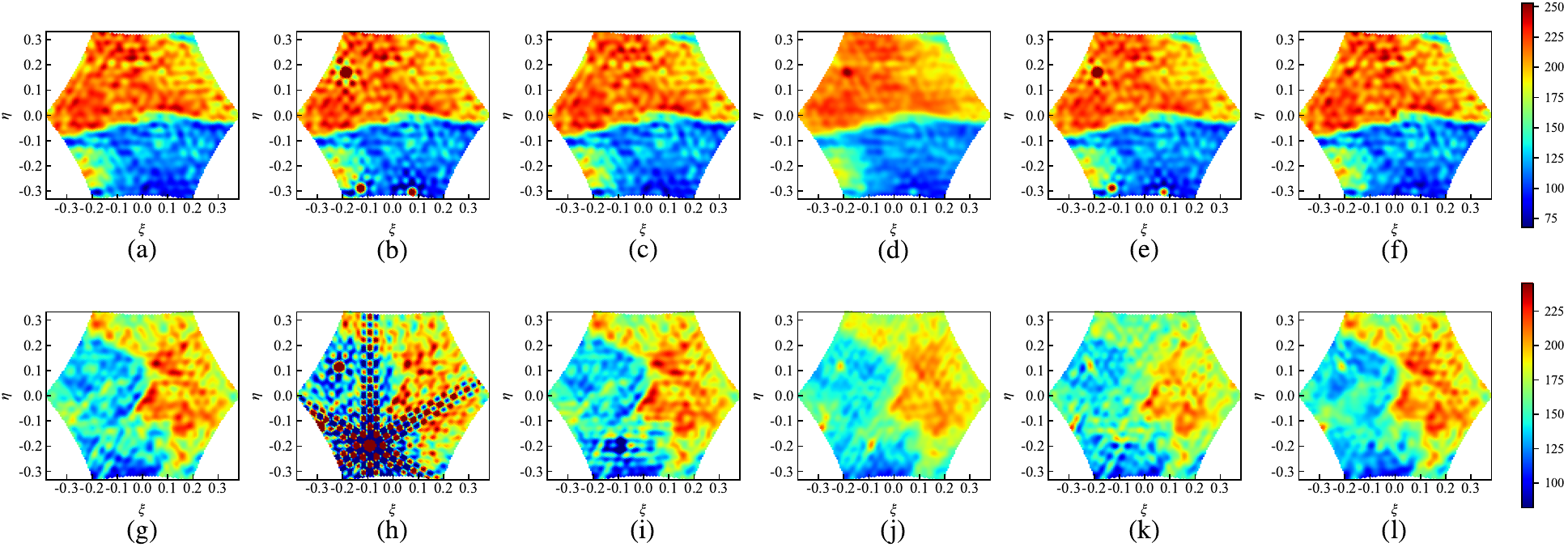}
    \caption{RFI mitigation examples on simulated data. (a) Scene 1 (snapshot ID: 691841314); (b) weak RFI contamination of scene 1 and (c)-(f) corresponding results after applying CLEAN,
RPCA, RNN-DFT, VFDM respectively.  (g) Scene 2 (snapshot ID: 691834271); (h) hybrid RFI contamination of scene 2 and (i)-(l) corresponding results after applying CLEAN,
RPCA, RNN-DFT, VFDM respectively.}
    \label{fig:grsl_simulated}
\end{figure*}

\subsection{Evaluation Based on Simulated Data} 
In this subsection, we evaluate the RFI mitigation performance of different methods
under various interference scenarios using simulated data.
Two background scenes used for data synthesis are shown in
\figsref{fig:grsl_simulated}{(a)} and \justsub{fig:grsl_simulated}{g}.
Scene~1 contains three weak RFI sources, leading to the contaminated observation
shown in \figref{fig:grsl_simulated}{b}.
Scene~2 (\figref{fig:grsl_simulated}{h}) represents a hybrid scenario with one strong
RFI source causing severe distortion and two additional weak sources.

Visual comparisons for CLEAN, RPCA, and RNN-DFT are shown in \figsref{fig:grsl_simulated}{(c)–(e)} and \figsref{fig:grsl_simulated}{(i)–(k)}, with VFDM results in \figref{fig:grsl_simulated}{f} and \figref{fig:grsl_simulated}{l}. In the weak RFI regime, CLEAN suppresses dominant interference but sacrifices detail to some extent. RPCA causes significant detail loss, while RNN-DFT fails due to model mismatch. Conversely, VFDM effectively mitigates RFI while preserving high-frequency details. In the hybrid case, CLEAN leaves voids at strong RFI locations, and both RPCA and RNN-DFT exhibit texture degradation. VFDM, however, eliminates both strong and weak RFI while maintaining structural integrity. Table~\ref{tab:experiment_simulated} details the quantitative results under both weak and hybrid scenarios.

Table~\ref{tab:experiment_dataset} summarizes the quantitative performance across various simulated scenarios. Our proposed method consistently achieves superior performance over all baselines in terms of both RMSE and SSIM. These results are derived from a comprehensive statistical analysis conducted on a test set of 1,310 image pairs, representing a 10\% split relative to the 11,697 training pairs. Such robust evaluation across the entire testing distribution underscores the effectiveness and generalizability of our approach in RFI mitigation.

\begin{table}[t]
\centering
\caption{Quantitative evaluation on representative simulated scenes}
\label{tab:experiment_simulated}
\scriptsize
\renewcommand{\arraystretch}{0.9} 
\setlength{\tabcolsep}{3pt} 

\begin{tabular}{@{}l cc cc@{}}
\toprule
\multirow{2}{*}{\textbf{Method}} & 
\multicolumn{2}{c}{\textbf{Scene 1 Weak}} & 
\multicolumn{2}{c}{\textbf{Scene 2 Hybrid}} \\
\cmidrule(lr){2-3} \cmidrule(lr){4-5}

& \textbf{RMSE (K)} & \textbf{SSIM}
& \textbf{RMSE (K)} & \textbf{SSIM}\\
\midrule

CLEAN    & 4.1448          & 0.9895          & 10.0059         & 0.9497 \\
RPCA     & 10.3200         & 0.9742          & 12.3607         & 0.9018 \\
RNN-DFT  & 11.1426         & 0.9516          & 17.0378         & 0.7981 \\
VFDM     & \textbf{3.3145} & \textbf{0.9975} & \textbf{8.0854} & \textbf{0.9638} \\
\bottomrule
\end{tabular}
\end{table}

\begin{table}[t]
\centering
\caption{Quantitative evaluation under different RFI modes.}
\label{tab:experiment_dataset}

\scriptsize
\renewcommand{\arraystretch}{1.1}
\setlength{\tabcolsep}{2.5pt}

\begin{tabular}{@{}l cc c cc c cc@{}}
\toprule
\multirow{2}{*}{\textbf{Method}} & 
\multicolumn{2}{c}{\textbf{Weak}} && 
\multicolumn{2}{c}{\textbf{Medium}} && 
\multicolumn{2}{c}{\textbf{Strong}} \\
\cmidrule{2-3} \cmidrule{5-6} \cmidrule{8-9}
& \textbf{RMSE (K)} & \textbf{SSIM}&&  
  \textbf{RMSE (K)} & \textbf{SSIM}&&  
  \textbf{RMSE (K)} & \textbf{SSIM}\\
\midrule

CLEAN & 4.6372 & 0.9678 && 5.2606 & 0.9582 && 8.1993 & 0.8865 \\
RPCA  & 9.3312 & 0.8690 && 9.2859 & 0.8588 && 10.0402 & 0.8368 \\
RNN-DFT   & 19.1234 & 0.6587 && 8.9923 & 0.8869 && 14.3030 & 0.7577 \\
VFDM  & \textbf{3.9376} & \textbf{0.9771} && \textbf{4.1397} & \textbf{0.9724} && \textbf{6.3945} & \textbf{0.9269} \\

\midrule
\addlinespace[4pt] 

\multirow{2}{*}{\textbf{Method}} & 
\multicolumn{2}{c}{\textbf{Very Strong}} && 
\multicolumn{2}{c}{\textbf{Hybrid}} && 
\multicolumn{2}{c}{} \\ 
\cmidrule{2-3} \cmidrule{5-6}
& \textbf{RMSE (K)} & \textbf{SSIM}&& 
  \textbf{RMSE (K)} & \textbf{SSIM}&& 
  \multicolumn{2}{l}{} \\

CLEAN & 26.0537 & 0.3834 && 10.3824 & 0.8773 && \multicolumn{2}{l}{} \\
RPCA  & 16.5131 & 0.6781 && 10.5212 & 0.8473 && \multicolumn{2}{l}{} \\
RNN-DFT   & 24.1975 & 0.3909 && 16.9697 & 0.7067 && \multicolumn{2}{l}{} \\
VFDM  & \textbf{11.9316} & \textbf{0.7720} && \textbf{6.8621} & \textbf{0.9347} && \multicolumn{2}{l}{} \\
\bottomrule
\end{tabular}
\end{table}

\subsection{Evaluation Based on Real Data}
\begin{figure*}[t] 
    \centering
    \includegraphics[width=1\textwidth]{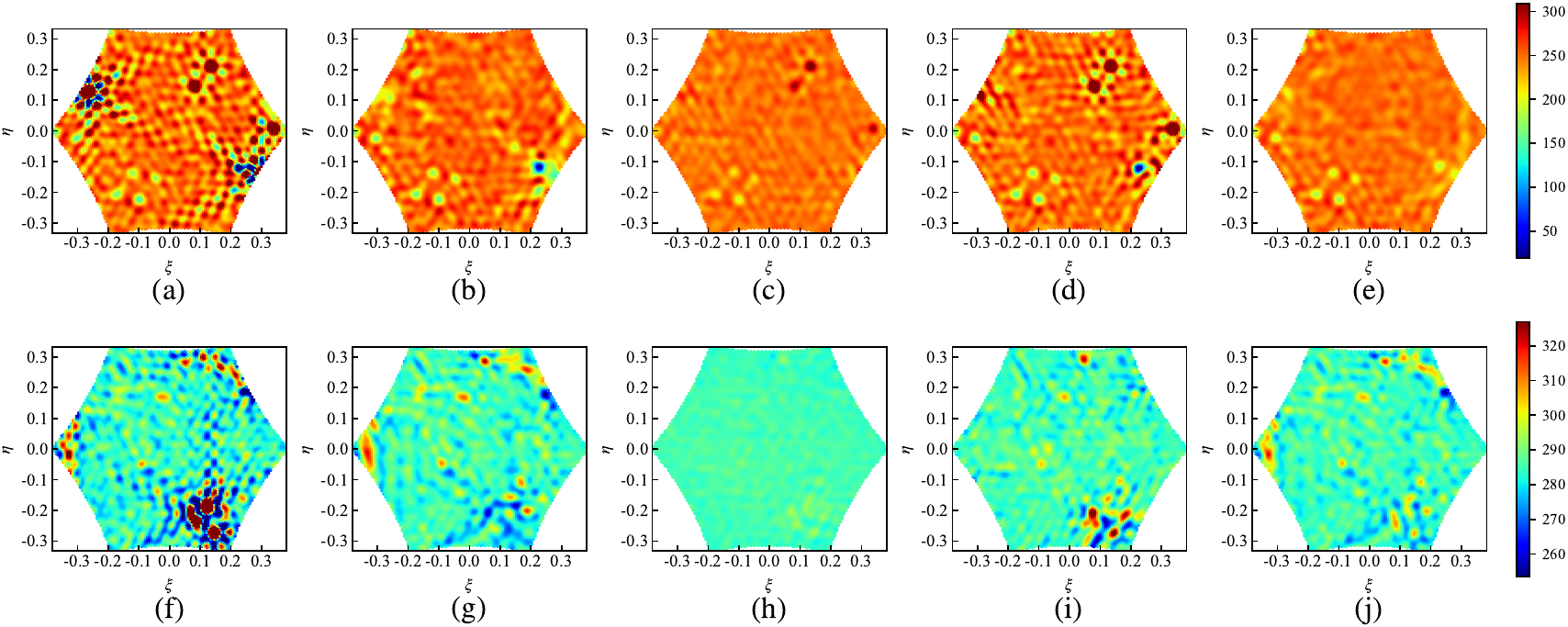}
    \caption{RFI mitigation examples on real data. (a) Scene 3 (snapshot ID: 691840761) and (b)-(e) corresponding results after applying CLEAN,
RPCA, RNN-DFT, VFDM respectively. (f) Scene 4 (snapshot ID: 691840910) and (g)-(j) corresponding results after applying CLEAN,
RPCA, RNN-DFT, VFDM respectively.}
    \label{fig:grsl_real} 
\end{figure*}

In this subsection, we evaluate the proposed method using SMOS L1A data: Scene 3 shown in \figref{fig:grsl_real}{a}, characterized by multiple RFI sources of varying intensities, and Scene 4 in \figref{fig:grsl_real}{f}, which features clustered RFI sources. Since clean reference BT images are unavailable for real-world observations, we adopt a reference-free metric, the TRE derived from the design philosophy in \cite{TRE}, to quantify mitigation performance:
\begin{equation}
\begin{split}
\text{TRE} = & \sqrt{\frac{1}{\|\mathbf{M}\|_0} \left\| \mathbf{M} \odot \left( T_{\text{dir}} - T_{\text{pred}}  \right) \right\|_F^2} \\
& + \frac{1}{2  \|1 - \mathbf{M}\|_0} \left\| (1 - \mathbf{M}) \odot \nabla T_{\text{pred}} \right\|_1.
\end{split}
\label{eq:TRE}
\end{equation}
where $\mathbf{M}$ denotes the binary mask, with values of $1$ indicating uncontaminated natural scene regions and $0$ indicating the presence of RFI. $T_{\text{dir}}$ and $T_{\text{pred}}$ are the dirty and predicted brightness temperatures, respectively, and $\nabla T_{\text{pred}}$ is the spatial gradient magnitude of the reconstruction. The symbol $\odot$ represents the Hadamard product.

The CLEAN method [\figref{fig:grsl_real}{b} and \justsub{fig:grsl_real}{g}] suppresses RFI in both scenarios but leaves noticeable signal voids at the locations of strong interference. RPCA [\figref{fig:grsl_real}{c} and \justsub{fig:grsl_real}{h}] fails to effectively remove weak RFI and suffers from detail loss in the hybrid case. Similarly, the RNN-DFT method [\figref{fig:grsl_real}{d} and \justsub{fig:grsl_real}{i}] cannot adequately suppress weak interference, introducing residual artifacts and voids. In contrast, the proposed VFDM [\figref{fig:grsl_real}{e} and \justsub{fig:grsl_real}{j}] effectively eliminates RFI while restoring a smoother and more physically consistent BT distribution.

Quantitative results in Table~\ref{tab:real} confirm this performance: VFDM achieves the lowest TRE scores (13.0893 for weak and 8.8218 for hybrid cases), validating its capability of recovering natural scene information.

\begin{table}[t]
\centering
\caption{Quantitative evaluation on real scenes (Total Reconstruction Error in K)}
\label{tab:real}
\scriptsize
\renewcommand{\arraystretch}{0.9}
\setlength{\tabcolsep}{6pt}

\begin{tabular}{@{}l cccc@{}}
\toprule
\textbf{Scene} 
& \textbf{CLEAN} 
& \textbf{RPCA} 
& \textbf{RNN-DFT} 
& \textbf{VFDM} \\
\midrule
Scene 3 
& 14.3045
& 15.1851
& 17.9321
& \textbf{13.0893} \\
Scene 4 
& 9.4263
& 9.8637
& 11.8205
& \textbf{8.8218} \\
\bottomrule
\end{tabular}
\end{table}

\section{Conclusion}
\label{section:conclusion}
This letter introduces the Visibility-Function Diffusion Model (VFDM) for RFI mitigation in SAIR, leveraging the generative capabilities of DDPMs within the spatial-frequency domain. To validate the model, we developed a physically-consistent simulation framework tailored to SAIR characteristics, alongside real-world SMOS data. Experimental results demonstrate that VFDM significantly outperforms the baselines. 
Future research will integrate spatial-frequency priors to further improve inference.

\bibliographystyle{IEEEtran}
\bibliography{refs}

\end{document}